# Nonlocal surface dipoles and vortices

Fangwei Ye, Yaroslav V. Kartashov, and Lluis Torner

*ICFO-Institut de Ciencies Fotoniques, and Universitat Politecnica de Catalunya,*

*Mediterranean Technology Park, 08860 Castelldefels (Barcelona), Spain*

We predict the existence and address the stability of two-dimensional surface solitons featuring topologically complex shapes, including dipoles, vortices, and bound states of vortex solitons, at the interface of nonlocal thermal media. Unlike their counterparts in bulk media, surface dipoles are found to be stable in the entire existence domain. Surface vortices are found to exhibit strongly asymmetric intensity and phase distributions, and are shown to be stable, too. Bound states of surface vortex solitons belong to a novel class of surface solitons having no counterparts in bulk media. Such states are found to be stable provided that their energy flow does not exceed an upper threshold. Our findings constitute the first known example of topologically complex solitons located at nonlocal two-dimensional interfaces.

Nonlocality of the nonlinear response is a generic property of a variety of nonlinear materials. Nonlocality arises when such nonlinearity mechanisms as diffusion of carriers, reorientation of molecules, heat transfer, etc., are involved [1]. Nonlocality may give rise to various nonlinear stationary light states [2-17] with no counterpart in local media. Especially rich situation is encountered in two-dimensional geometries, where materials with different types of nonlocal response can support stationary multipoles [7-11], stable vortices [12-16], rotating [17,18] and spiraling [19,20] soliton states. Thus, thermal nonlinearities arising due to heat transfer phenomena [15] may result in stabilization of ring-shaped vortices with charges one and two [15,16]. Thermal media have been also utilized to demonstrate stationary multipole-mode solitons [7]. Such multipoles are weakly unstable, although the possibility of their stabilization by setting them into rotation was suggested [17]. Importantly, note that all types of solitons mentioned above have been studied in bulk media. Nevertheless, the presence of interfaces in the nonlocal medium may substantially affect soliton properties and conditions required for their excitation. Under appropriate conditions light can attach to the interface resulting in the formation of specific surface waves.

Stationary surface waves propagating along the interface of two different optical materials exhibit unique properties, which have been comprehensively studied in local media. Surface waves in nonlocal media have been addressed only recently. An important result that has been uncovered is that the geometry of nonlocal sample may play a key role in the surface wave existence [15,21]. Surface waves at the interface of two Kerr-type materials, characterized by a finite degree of nonlocality, were studied in Refs. [22,23]. In focusing thermal media where the nonlocality range is determined by the transverse extent of the very thermal sample, two-dimensional fundamental surface waves were observed in Ref. [24]. Defocusing thermal materials can also support surface waves under appropriate conditions [25]. However, to date no complex surface solitons with rich internal structure have been reported in two-dimensional nonlocal materials. Therefore, a fundamental question arises: Which types of higher-order surface states could exist at interfaces of two-dimensional nonlocal media, and how their properties would differ from the properties of higher-order solitons in uniform nonlocal media?

In this paper we predict the existence and study the stability of different soliton states with nontrivial topological structures supported by the interface of thermal medium. This includes sur-face dipoles, vortices, as well as bound states of surface vortex soliton. We find that surface dipoles are stable in their entire existence domain, in clear contrast to dipoles in uniform thermal medium. Surface vortices feature strongly anisotropic and noncanonical shapes and can also be completely stable. Nonlocal thermal interface can support a novel stationary bound state of vortices with no counterpart in uniform media.

We start our analysis by considering a laser beam propagating along the $\xi$ axis in the vicinity of the interface formed by the nonlocal thermal medium and a linear medium. The propagation of laser beam is described by the system of equations for dimensionless complex field amplitude $q$ and nonlinear contribution to the refractive index $n$ (see, e.g., Ref. [24] for a detailed description of the model):

$$\begin{cases} i\dfrac{\partial q}{\partial \xi} = -\dfrac{1}{2}\Delta_\perp q - nq \\ \Delta_\perp n = -|q|^2 \end{cases} \text{ in thermal medium}$$
$$\begin{cases} i\dfrac{\partial q}{\partial \xi} = -\dfrac{1}{2}\Delta_\perp q - n_d q \quad \text{in linear medium} \end{cases} \tag{1}$$

Here $q = (k_0^2 r_0^4 \alpha\beta / \kappa n_0)^{1/2} A$ is the dimensionless light field amplitude; $n = k_0^2 r_0^2 \delta n / n_0$ is proportional to the nonlinear change $\delta n$ in the refractive index $n_0$ (note that the nonlinear refractive index change is given by $\delta n = \beta T$, where $T$ is the temperature variation, that obeys the Laplace equation $\kappa\Delta_\perp T = \alpha I$); $\alpha, \beta, \kappa$ are the optical absorption, thermo-optic, and thermal conductivity coefficients, respectively; $I$ is the light intensity; $k_0$ is the wavenumber; $\Delta_\perp = \partial^2/\partial\eta^2 + \partial^2/\partial\zeta^2$ is the transverse Laplacian; the transverse $\eta, \zeta$ and longitudinal $\xi$ coordinates are scaled to charac-teristic beam width $r_0$ and diffraction length $k_0 r_0^2$, respectively; the parameter $n_d < 0$ describes the difference between the unperturbed refractive index $n_0$ of thermal medium and refractive index of less optically dense linear medium. We assume that the thermal medium occupies the spatial region $-L/2 \leq \eta \leq 0$ and $-L/2 \leq \zeta \leq L/2$, where $L$ is the $\zeta$-width of the sample.

Nonlocal thermal responses are often described as exhibiting an infinite range of the nonlocality, because the conditions imposed at the boundaries of the sample greatly affect the entire refractive index distribution. This effect enters the model via the second of Eqs. (1), which is a Laplace equation for the nonlinear refractive shift, with a source term proportional to the light intensity. The solution of this equation strongly depends on the boundary conditions. Here we assume that the thermal medium is in contact with a linear medium at $\eta = 0$ and that the thermal conductivity $\kappa_d$ of the linear medium is much smaller than that of the thermal medium. Physically, this means that the interface at $\eta = 0$ is thermally insulating and that the thermal flux through this interface is negligible, i.e., vanishing $\partial n(\eta = 0)/\partial \eta = 0$. Three other boundaries $\eta = -L/2$, $\zeta = \pm L/2$ of the thermal sample are maintained at the same temperature so that one can assume that $n = 0$ at these boundaries. Note that, formally, in the model considered here the refractive index can always be set to zero because adding a constant background in the refractive index is equivalent to introducing a shift of the soliton propagation constant. We set $L = 40$, $n_d = -100$, which correspond to typical experimental conditions. Note that when $|n_d| \gg 1$ the surface waves residing in the vicinity of $\eta = 0$ interface penetrate into the linear medium only slightly. The system (1) conserves the total energy flow

$$U = \iint_{-\infty}^{\infty} |q|^2 \, d\eta d\zeta. \tag{2}$$

On physical grounds, a laser beam launched in the vicinity of the interface of the thermal medium experiences slight absorption upon propagation and thus raises the temperature of the surrounding material. Due to diffusion of heat that occurs predominantly in the direction of the insulating surface, a thermal lens forms inside the medium as a result of the thermo-optic effect (see, e.g., Ref. [24]). This thermal lens results in the deflection of the laser beam towards the insulating surface, a phenomenon that under appropriate conditions may result in the formation of stationary surface wave.

We searched for surface solitons of Eq. (1) that exhibit topologically complex internal structures, such as dipole solitons that have the form $q(\eta,\zeta,\xi) = w\exp(ib\xi)$, and vortex solitons whose field can be written in the form $q(\eta,\zeta,\xi) = (w_\mathrm{r} + iw_\mathrm{i})\exp(ib\xi)$. Here $w(\eta,\zeta)$ and $w_\mathrm{r,i}(\eta,\zeta)$ are real functions independent of the propagation distance $\xi$, while $b$ is the propagation constant. Substitution into the second of Eqs. (1) leads to the two-dimensional nonlinear refractive index profile $n(\eta,\zeta)$ corresponding to the stationary solution. The surface solutions were found numerically by a standard relaxation method. In all cases analyzed in this paper, the method converged to a stationary solution after several iterations provided that a suitable initial guess for field distribution (real in the case of dipoles or complex in the case of vortices) and refractive index are selected. In all cases the continuity conditions $w(\eta,\zeta)|_{\eta \to +0} = w(\eta,\zeta)|_{\eta \to -0}$, $\partial w(\eta,\zeta)/\partial \eta|_{\eta \to +0} = \partial w(\eta,\zeta)/\partial \eta|_{\eta \to -0}$ are satisfied at the interface at $\eta = 0$. The obtained surface solitons are characterized by complex topological internal structures. Thus, dipole solutions consist of two out-of-phase bright spots with a nodal (zero-intensity) line separating them. We have found two different types of dipole solitons residing in the vicinity of the interface: the nodal lines in solitons of the first type are almost parallel to the interface, while in solitons of the second type they are perpendicular to the interface. Here we concentrate on soliton solutions of the second type because of their enhanced stability. The amplitude, phase, and induced refractive index profile for dipole soliton are depicted in Fig. 1. One can resolve two maxima in the refractive index distribution whose positions approximately correspond to peaks in the intensity distribution, but it should be pointed out that due to strongly nonlocal character of the thermal response, the nonlinear refractive index does not vanish even in the zero-intensity regions between bright spots forming dipole; rather, it is only slightly reduced there. The refractive index distribution width largely exceeds that of solitons. Thus, even though they are located in close proximity to the interface, the dipole solitons modify the refractive index in the entire thermal medium [Fig. 1(c)]. The poles of solitons are slightly asymmetric in $\eta$ direction.

We also found a family of surface vortex solitons [Figs. 2(a)-(c)]. Such states carry phase singularity where light intensity vanishes. In the vortex solution depicted in Fig. 2, the phase increases by $2\pi$ upon winding around the singularity along any closed contour, so that the topological charge of such vortex equals to one. In contrast to their counterparts in uniform media, surface vortex solitons feature noncanonical intensity and phase distributions [26]. Thus, two maxima located on $\eta$-axis are clearly resolvable in the intensity distribution, while the maximum located farther from the interface is more pronounced. The vortex intensity modulation becomes more pronounced with increasing energy flow. Unlike in canonical radially symmetric vortices featuring constant phase gradients $d\theta/d\phi$ (here $\theta$ is the vortex phase and $\phi$ is azimuthal angle), for our surface vortices $d\theta/d\phi$ is azimuthally dependent [Fig. 4(c)]. The phase increases most rapidly around $\phi = \pi/2$ and $3\pi/2$, while $d\theta/d\phi$ is minimal in the vicinity of the intensity maxima. Like dipole solitons, surface vortices modify the refractive index distribution in the entire sample [Fig. 2(c)].

In addition to surface dipoles and vortices, we found a novel type of stationary surface soliton in the form of a bound state of two vortices (see Fig. 3). Such bound states were also found in the form $q(\eta,\zeta,\xi) = (w_r + iw_i)\exp(ib\xi)$. To the best of our knowledge, such stationary nonrotating bound states have not been encountered previously in bulk nonlocal materials. Therefore, the surface geometry appears to be necessary for their existence. One can clearly see from Fig. 3(b) that vortices forming the bound state have the same topological charge. The vortex located closer to the interface experiences much stronger shape deformation. The refractive index distribution is more elongated along the $\eta$-axis for bound state of vortices than for dipole and usual vortex solitons [this is readily visible in Fig. 3(c)]. The intensity maximum located farther from the interface is always more pronounced than other two maxima lying on $\eta$-axis.

Figure 4(a) shows the energy flow as a function of the propagation constant for dipole surface solitons. The corresponding dependence for surface vortices is quite similar and therefore we do not show it here. One can see that for both surface dipoles and vortices the energy flow is a monotonically increasing function of the propagation constant. With increasing energy flow, surface solitons become more localized: for dipoles the distance between poles along $\eta$-axis and their widths decrease, while for vortices the integral width of the total complex intensity distribution also decreases. The distance between the vortex phase singularity and the interface gradually decreases with $U$. For solitons of all types discussed here, the energy flow vanishes when $b \to 0$.

To elucidate the dynamical stability of the complex surface solitons, we performed comprehensive simulations of propagation of surface states perturbed by input noise with variance $\sigma_{\text{noise}}^2 = 0.01$. Importantly, we found that surface dipoles are stable in the entire existence domain. A representative example of the typical evolution dynamics is shown in Fig. 5(a). The considerable input perturbations cause only small oscillations of amplitudes of the two poles forming the dipole, but the dipoles preserve their internal structures over huge distances, exceeding the experimentally available crystal lengths by several orders of magnitude. This is in contrast to dipoles in bulk thermal medium, where weak input perturbations cause slow but progressively increasing oscillations of bright spots forming dipole resulting typically in their slow decay into fundamental solitons. Therefore, we conclude that the presence of thermally insulating interface results in stabilization of dipole solitons.

Surface vortices featuring strongly asymmetric and noncanonical shapes are also found to be completely stable in the entire existence domain, in analogy with their radially symmetric counterparts in bulk geometries [see Fig. 5(b) showing the stable propagation of a perturbed surface vortex soliton]. Since boundary effects play a crucial role in the properties of solitons in thermal medium, it is important to elucidate whether the aforementioned stability of complex surface solitons (dipoles, in particular) is associated with surface effects. We thus varied the transverse size of thermal medium and the ratio of its $\eta$ and $\zeta$ dimensions (recall, that this ratio was 1⁄2 in all previous simulations, i.e. the sample was rectangular) and tested the stability of dipole and vortex solitons in each case. Our extensive simulations confirmed that both dipole and vortex solitons remain stable in their entire existence domain, irrespectively of the overall geometry of the sample.

Finally, strongly asymmetric bound states of vortex solitons were also found to be stable [see an example of stable evolution in Fig. 5(c)], provided that the energy flow does not exceed a certain critical value. The stable and unstable branches for vortex bound states are depicted in Fig.

4(b), showing the $U(b)$ dependence with continuous and dashed lines, respectively. We found that the width of the stability domain for such bound states strongly depends on the overall geometry of the thermal sample (i.e., the aspect ratio, as described above).

With a laser at the wavelength $\sim 500$ nm and beam width $\sim 50\,\mu$m, observation of the surface states reported above requires nonlinear contributions to the refractive index of the order of $\delta n \sim 10^{-5}$. In a typical thermal medium, such as lead glass, with $n_0 = 1.8$, $\beta \sim 10^{-5}\,\text{K}^{-1}$, $\alpha \sim 0.01\,\text{cm}^{-1}$, and $\kappa \sim 1\,\text{Wm}^{-1}\text{K}^{-1}$ [15,24], such nonlinear contributions are readily achievable with input light powers of the order of $1\,\text{W}$. With these parameters $n_d = -100$ corresponds to a refractive index difference between the thermal medium and the surrounding linear material of about $\sim 10^{-4}$. Notice that this small difference in linear indices justifies the assumption of equal diffraction coefficients in the thermal and linear media in Eq. (1), since the relative difference in the diffraction coefficients is of the same order as the refractive index difference. We emphasize that our findings remain valid as long as $|n_d| \gg 1$, since in this case surface waves almost do not penetrate into the linear material and almost all power is concentrated within the thermal medium. Also, increasing $|n_d|$ does not change this picture.

Summarizing, we predicted the existence of a variety of surface solitons with nontrivial topological structures at thermally insulating interfaces between thermal and linear media. We found that surface dipoles and vortices are always stable at such interfaces, while complex bound states of vortex solitons can be stable below certain energy threshold. We also uncovered the existence of bound states of surface vortex solitons with no counterparts in bulk media. Our findings constitute the first known example of stable topologically complex surface solitons supported by two-dimensional interfaces located in nonlocal media.

This work has been partially supported by the Government of Spain through grant TEC2005-07815 and by the Ramon-y-Cajal program.

# Figure captions

Figure 1 (color online).   (a) Field modulus, (b) phase, and (c) refractive index profile for surface dipole at $b=3$. White dashed line indicates interface position. All quantities are plotted in arbitrary dimensionless units.

Figure 2 (color online).   (a) Field modulus, (b) phase, and (c) refractive index profile for surface vortex at $b=3$. All quantities are plotted in arbitrary dimensionless units.

Figure 3 (color online).   (a) Field modulus, (b) phase, and (c) refractive index profile for bound state of surface vortex solitons at $b=3$. All quantities are plotted in arbitrary dimensionless units.

Figure 4 (color online).   Energy flow versus propagation constant for (a) surface dipole and (b) bound state of vortex solitons. Points marked with solid-line circles correspond to solitons shown in Figs. 1 and 3. Continuous curve corresponds to stable branch, dashed curve corresponds to unstable branch. (c) Noncanonical phase distribution for vortex surface soliton defined on a closed contour surrounding phase singularity at $b=3$. All quantities are plotted in arbitrary dimensionless units.

Figure 5 (color online). Stable propagation of (a) surface dipole, (b) vortex, and (c) bound state of vortex solitons with $b=3$. In all cases white noise with variance $\sigma_{\text{noise}}^{2}=0.01$ was added into input distributions. Field modulus distributions are shown at different propagation distances. All quantities are plotted in arbitrary dimensionless units.

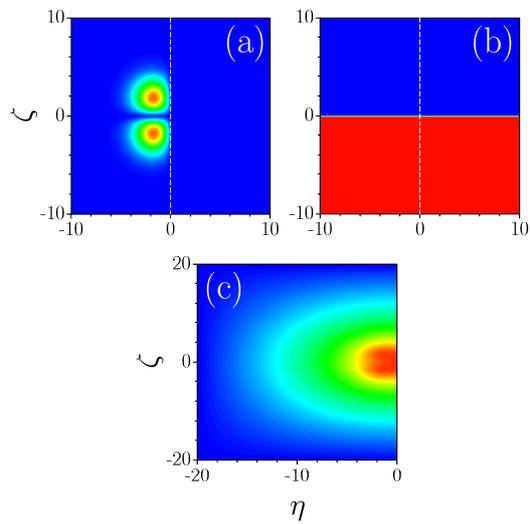

Figure 1 (color online). (a) Field modulus, (b) phase, and (c) refractive index profile for surface dipole at $b=3$. White dashed line indicates interface position. All quantities are plotted in arbitrary dimensionless units.

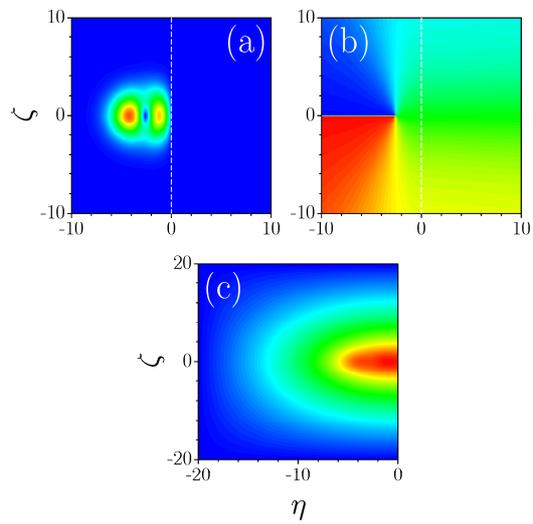

Figure 2 (color online). (a) Field modulus, (b) phase, and (c) refractive index profile for surface vortex at $b=3$. All quantities are plotted in arbitrary dimensionless units.

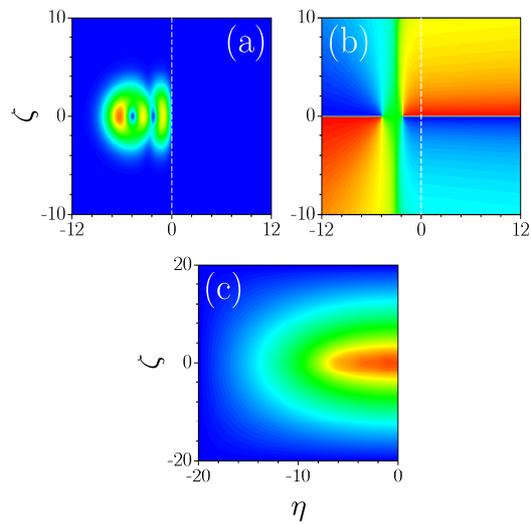

Figure 3 (color online). (a) Field modulus, (b) phase, and (c) refractive index profile for bound state of surface vortex solitons at $b=3$. All quantities are plotted in arbitrary dimensionless units.

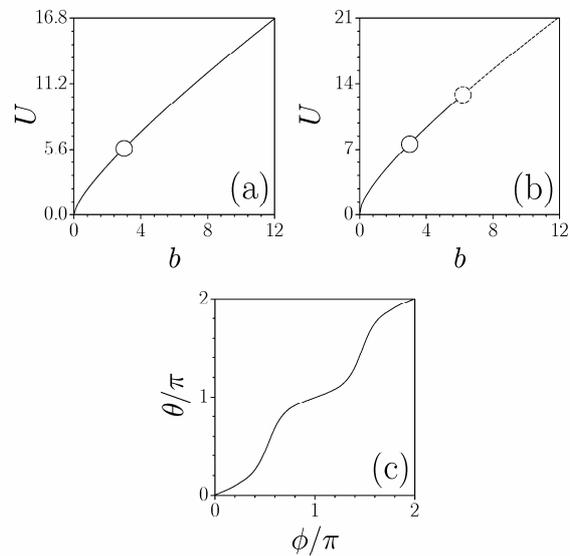

Figure 4 (color online). Energy flow versus propagation constant for (a) surface dipole and (b) bound state of vortex solitons. Points marked with solid-line circles correspond to solitons shown in Figs. 1 and 3. Continuous curve corresponds to stable branch, dashed curve corresponds to unstable branch. (c) Noncanonical phase distribution for vortex surface soliton defined on a closed contour surrounding phase singularity at $b=3$. All quantities are plotted in arbitrary dimensionless units.

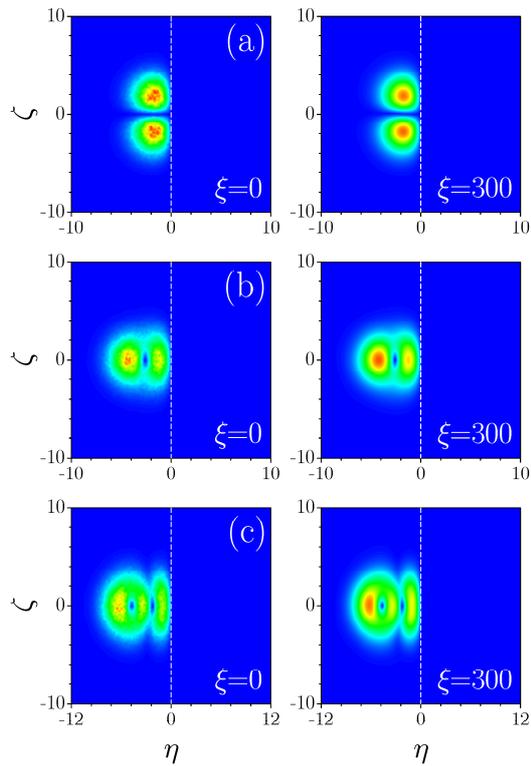

Figure 5 (color online). Stable propagation of (a) surface dipole, (b) vortex, and (c) bound state of vortex solitons with $b=3$. In all cases white noise with variance $\sigma_{\text{noise}}^2 = 0.01$ was added into input distributions. Field modulus distributions are shown at different propagation distances. All quantities are plotted in arbitrary dimensionless units.

20